\documentstyle[12pt]{article}

\begin{document}


\def\a{\alpha}
\def\b{\beta}
\def\c{\varepsilon}
\def\d{\delta}
\def\e{\epsilon}
\def\f{\phi}
\def\g{\gamma}
\def\h{\theta}
\def\k{\kappa}
\def\l{\lambda}
\def\m{\mu}
\def\n{\nu}
\def\p{\psi}
\def\q{\partial}
\def\r{\rho}
\def\s{\sigma}
\def\t{\tau}
\def\u{\upsilon}
\def\v{\varphi}
\def\w{\omega}
\def\x{\xi}
\def\y{\eta}
\def\z{\zeta}
\def\D{\Delta}
\def\G{\Gamma}
\def\L{\Lambda}
\def\F{\Phi}
\def\P{\Psi}
\def\S{\Sigma}

\def\o{\over}

\def\beq{\begin{equation}}
\def\eeq{\end{equation}}

\def\NPB#1#2#3{{Nucl.~Phys.} {\bf{B#1}} (19#2) #3}
\def\PLB#1#2#3{{Phys.~Lett.} {\bf{B#1}} (19#2) #3}
\def\PRD#1#2#3{{Phys.~Rev.} {\bf{D#1}} (19#2) #3}
\def\PRL#1#2#3{{Phys.~Rev.~Lett.} {\bf#1} (19#2) #3}
\def\PTP#1#2#3{{Prog.~Theor.~Phys.} {\bf#1} (19#2) #3}

\newcommand{\siml}{\raise -2.truept\hbox{\rlap{\hbox{$\sim$}}\raise5.truept
\hbox{$<$}\ }}
\newcommand{\simg}{\raise -2.truept\hbox{\rlap{\hbox{$\sim$}}\raise5.truept
\hbox{$>$}\ }}

\begin{titlepage}
\vspace{3cm}
\begin{flushright}
UT-812 \\
TU-544, RCNS-98-06 \\

April 1998
\end{flushright}
\vskip 0.8cm
\begin{center}
{\Large \bf Gravitino Overproduction \\ through Moduli Decay}
\end{center}
\vskip 1.2cm
\begin{center}
M. Hashimoto$^{1}$\renewcommand{\thefootnote}{\fnsymbol{footnote}}%
\footnote{Research Fellow of the Japan Society for the Promotion of
Science.},
Izawa K.-I.$^1$, M. Yamaguchi$^2$ and T. Yanagida$^1$
\\
\vskip 1.5cm
{\em $^1$Department of Physics, University of Tokyo, \\
Tokyo 113-0033, Japan}\\ 
\vskip 1cm
{\em $^2$Department of Physics, Tohoku University, \\
Sendai 980-8578, Japan}
\end{center}
\vskip 2cm
\thispagestyle{empty}
\begin{abstract}
We derive cosmological constraints on the masses of generic scalar fields
which decay only through gravitationally suppressed interactions
into unstable gravitinos and ordinary particles in the supersymmetric
standard model.
For the gravitino mass 100GeV-1TeV, the scalar masses should be larger than
100TeV to keep the success of big-bang nucleosynthesis
if no late-time entropy production dilutes the gravitino density. 
\end{abstract}
\end{titlepage}


\section{Introduction}

Superstring theories
have infinitely degenerate supersymmetric vacua which are continuously
connected by massless scalar fields, called moduli.
These moduli fields are generally expected to acquire their masses
$m_\v$ of the order of the gravitino mass $m_{3/2}$ once
supersymmetry breaking effects are included.
The moduli decay into two gravitinos
if the masses of moduli are larger than $2m_{3/2}$.

In this paper, we consider the decay of the moduli into two gravitinos
and discuss its cosmological consequences.
It is known
\cite{Wei,Kaw}
that the gravitino with the mass 100GeV-1TeV decays soon after
the nucleosynthesis and the decay product destroys light nuclei
produced in the early universe.
We see that the moduli decay tends to produce too many gravitinos
to keep the success of big-bang nucleosynthesis.
We derive stringent constraints on the moduli masses
to avoid this disaster such as $m_\v \simg 100{\rm TeV}$.

We stress that this constraint is applicable to generic scalar fields
that decay only through gravitationally suppressed interactions
as long as their masses are larger than the threshold of two-gravitino
decay channel.

\section{The Interaction}

We assume one modulus field throughout the paper,
though the generalization to the case of many moduli is straightforward.
We set the gravitational scale $2.4 \times 10^{18}$GeV equal to unity.

The relevant terms in the supergravity Lagrangian
\cite{Wes}
which describe
the decay of a modulus $\v$ into gravitinos $\chi$ is given by
\beq
 {\cal L} = \e^{klmn}{\bar \chi}_k{\bar \s}_l
             {1 \o4}(K_\v \q_m\v-K_{\v^*}\q_m\v^{*})\chi_n
          - e^{K \o 2}(W^*\chi_a\s^{ab}\chi_b + W{\bar \chi}_a
            {\bar \s}^{ab}{\bar \chi}_b),
\label{LAG}
\eeq
where $W$ denotes the superpotential and we choose the field $\v$
so that its vacuum expectation value
vanishes: $\langle \v \rangle = 0$. With our definition of $\v$,
the K\"ahler potential $K$ generically contains linear terms,
$K = c\v + c^*\v^* + \v \v^* + \cdots$, where the coefficient $c$ is
of order one.

\section{The Decay}

Let us begin by discussing the decay rate of the modulus.
When $m_\v \gg 2m_{3/2}$,
the order of the decay width of $\v$ into gravitinos is given by
\beq
 \G(\v \rightarrow {\rm gravitinos}) \sim |c|^2m_{3/2}^2m_\v,
\eeq
where $c$ denotes the coefficient of the $\v$ term
in the K\"ahler potential.
Here, we have used Eq.(\ref{LAG}), the gravitino equation of motion
and $\langle e^{K/2}W \rangle = m_{3/2}$.

On the other hand, the order of the decay width of $\v$
into radiation is given by
\beq
 \G(\v \rightarrow {\rm radiation}) \sim Nm_\v^3,
\eeq
where $N$ is the number of the decay channels.
Hence the branching ratio of the decay into gravitinos
turns out to be
\beq
 B_\chi \sim {|c|^2 \o N}\left( {m_{3/2} \o m_\v} \right)^2.
\eeq

The modulus $\v$ starts damped oscillation when the Hubble scale $H$
becomes
comparable to its mass $m_\v$.
The initial amplitude of the coherent oscillation is expected
to be of order one in the Planck unit.
Then the modulus density $\r_\v$ dominates
the universe at the decay time since $\G_\v \ll H \sim m_\v$.
The reheat temperature after the modulus decay is given by
\beq
 T_R \sim N_*^{-{1 \o 4}}\sqrt{\G_\v},
\eeq
where $N_*$ denotes the degrees of freedom at the temperature $T_R$.

This implies that the gravitino number density $n_{3/2}$ produced through
the modulus decay at the decay time is given by
\beq
 {n_{3/2} \o s} \sim {|c|^2N_*^{-{1 \o 4}} \o \sqrt{N}}
                     {m_{3/2}^2 \o {m_\v}^{3 \o 2}},
\eeq
where $s$ denotes the entropy density and we have used
\beq
 B_\chi \r_\v \sim m_{\v}n_{3/2}, \quad \r_\v \sim N_* T_R^4, \quad
 s \sim N_* T_R^3.
\eeq
Namely, the modulus mass is given by
\beq
 {m_\v} \sim \left( {|c|^2N_*^{-{1 \o 4}} \o \sqrt{N}}
                    {m_{3/2}^2 \o y_{3/2}} \right)^{2 \o 3},
\label{MAS}
\eeq
where $y_{3/2} = n_{3/2}/s$.

Gravitinos are also produced by the scattering processes
of the thermal radiations after the modulus decay.
The contribution to the gravitino number density is given by
\cite{Kaw}
\beq
 {n'_{3/2} \o s} \sim 10^{-3} T_R
                 \sim 10^{-3} N_*^{-{1 \o 4}} \sqrt{N} m_\v^{3 \o 2}.
\label{THE}
\eeq

\section{The Bound}

In the previous section, we have estimated $y_{3/2} = n_{3/2}/s$
at the decay time of the modulus $\v$.
We may derive cosmological constraints on $y_{3/2}$ from the observation
of the present universe since
the estimated value $y_{3/2}$ itself yields the value of $n_{3/2}/s$
at the time of the gravitino decay.

Stringent constraints
are implemented to keep the successful
predictions of the big-bang nucleosynthesis
provided the gravitino with the mass 100GeV-1TeV
mainly decays into a photon and a photino
\cite{Kaw}:
\beq
 y_{3/2} \siml 10^{-15}-10^{-13}.
\eeq
By means of Eq.(\ref{MAS}), we obtain
\beq
 m_\v \simg 100 {\rm TeV}.
\eeq

On the other hand, the constraints due to
gravitinos produced by the scattering processes
of the thermal radiations read as follows:
\beq
 {m_\v} \siml (10^{7}-10^{9}){\rm TeV},
\eeq
where we have used Eq.(\ref{THE}).

\section{Conclusion}

We have derived the constraint $100{\rm TeV} \siml m_\v \siml 10^9{\rm
TeV}$
on the moduli masses $m_\v$ in the case of the unstable gravitino%
\footnote{On the other hand, no stringent constraint on the moduli masses
is implemented due to the moduli decay in the case of the lighter gravitino
$m_{3/2} \siml 10{\rm GeV}$
\cite{Pag}.}
with the mass 100GeV-1TeV.
This may have obvious implications for mechanisms of the moduli
stabilization.
Here, the moduli may be regarded as generic scalar fields
that decay only through gravitationally suppressed interactions
as long as their masses are larger than the threshold of two-gravitino
decay channel.

In the course of the analysis, we assumed that
no entropy production has diluted the modulus and gravitino densities
since the modulus density once dominated the universe.
In fact, entropy production may evade the constraints.
New inflation and thermal inflation are possible candidates of
enough entropy production.
Without such inflationary dilution, the moduli masses are severely
constrained.
On the other hand, if the moduli masses lie in the region
$100 {\rm TeV} \siml m_\v \siml 10^9 {\rm TeV}$,
the reheat temperature of the cosmological inflation
could be very high since the density of gravitinos produced
just after the inflation is diluted substantially by the decay of moduli.

\section*{Acknowledgements}

The work of M.Y.~was supported in part by 
the Grant--in--Aid for Scientific Research from the Ministry of 
Education, Science and Culture of Japan No.\ 09640333.

\end{document}